# Manuscript Details

| | |
|---|---|
| **Manuscript number** | JOT_2016_3 |
| **Title** | Lumbar degenerative spondylolisthesis epidemiology: a systemic review with a focus on gender-specific and age-specific prevalence |
| **Short title** | Lumbar degenerative spondylolisthesis epidemiology |
| **Article type** | Review Article |

## Abstract

The epidemiology of lumbar degenerative spondylolisthesis (DS) remains controversial. We performed a systemic review with the aim to have a better understanding of DS's prevalence in general population. The results showed the prevalence of DS is very gender specific and age specific. Both women and men have few DS before 50 years old, after 50 years old both women and men start to develop DS, with women having a faster developing rate than men. For elderly Chinese (≥ 65 yrs, mean age: 72.5 yrs), large population based studies (MsOS(Hong Kong) and MrOS (Hong Kong), females n=2000 and males n=2000) showed DS prevalence was for 25.0% for women and 19.1% for men, and the prevalence F:M (women:men) ratio was 1.3:1. The published data (MsOS(USA) and MrOS(USA) studies) seem to show elderly Caucasian American has a higher DS prevalence, being approximately 60-70% higher than elderly Chinese; however the prevalence F:M ratio was similar to elderly Chinese population. Patient data showed female patients more often received treatment than men; and preliminary data show the ratio of numbers of female patients received treatment compared with men did not differ between Northeast Asians (Chinese, Japanese, and Korean) and European and American Caucasians, being around 2:1 in elderly population. The existing data also suggest that menopause may be a contributing factor for the accelerated DS development in post-menopausal women.

| | |
|---|---|
| **Keywords** | Degenerative spondylolisthesis; Prevalence; Chines;, Caucasian; Men; Women. |
| **Taxonomy** | Geographic Location, Patient Characteristics, Anatomy, Clinical Anatomy |
| **Manuscript category** | Rehabilitation |
| **Corresponding Author** | Yi-Xiang Wang |
| **Order of Authors** | Yi-Xiang Wang, Zoltán Káplár, Min Deng, Jason Leung |

## Submission Files Included in this PDF

**File Name [File Type]**

CoverLetter.docx [Cover Letter]

No_Conflict.docx [Conflict of Interest]

Spondylolysis_MS V3NewNew.doc [Manuscript (without Author Details)]

Fig_1.tif [Figure]

Fig_2.tif [Figure]

Fig_3.tif [Figure]

Fig_4.tif [Figure]

Fig_5.tif [Figure]

Fig_6.tif [Figure]

Table 1.docx [Table]

Lumbar degenerative spondylolisthesis epidemiology_titlePage.docx [Title Page (with Author Details)]

To view all the submission files, including those not included in the PDF, click on the manuscript title on your EVISE Homepage, then click 'Download zip file'.

Prof Ling Qin,

Editor-in-Chief, JOT.

Dear Professor Qin,

We would like to submit a systemic review paper entitled '*Lumbar degenerative spondylolisthesis epidemiology: a systemic review with a focus on gender-specific and age-specific prevalence*' for considerations of publication in *JOT*.

The prevalence, female male prevalence ratio, as well as risk factors of remain degenerative spondylolisthesis remains controversial. Our systemic review nicely reconciled the literatures, and pointed out degenerative spondylolisthesis prevalence is very age-specific and gender-specific. Our comparative analysis showed the difference of degenerative spondylolisthesis prevalence between males and females, and between Asians and Caucasians, is smaller than many papers suggested.

We feel this paper is very important and will represent a milestone for Lumbar degenerative spondylolisthesis epidemiology.

This paper is not under consideration in anywhere else.

Yours

Wang Yi-Xiang
The Chinese University of Hong Kong

Prof Ling Qin,

Editor-in-Chief, JOT.

Dear Professor Qin,

We would like to submit a systemic review paper entitled '*Lumbar degenerative spondylolisthesis epidemiology: a systemic review with a focus on gender-specific and age-specific prevalence*' for considerations of publication in *JOT*.

The prevalence, female male prevalence ratio, as well as risk factors of remain degenerative spondylolisthesis remain controversial. Our systemic review nicely reconciled the literatures, and pointed out degenerative spondylolisthesis prevalence is very age-specific and gender-specific. Our comparative analysis showed the difference of degenerative spondylolisthesis prevalence between males and females, and between Asians and Caucasians, is smaller than many papers suggested.

This paper is not under consideration in anywhere else.

We declare that there is no conflict of interests for this paper.

Yours

Wang Yi-Xiang

The Chinese University of Hong Kong

Article type: review (systemic review)

Lumbar degenerative spondylolisthesis epidemiology: a systemic review with a focus on gender-specific and age-specific prevalence

Abstract

The epidemiology of lumbar degenerative spondylolisthesis (DS) remains controversial. We performed a systemic review with the aim to have a better understanding of DS's prevalence in general population. The results showed the prevalence of DS is very gender-specific and age-specific. Both women and men have few DS before 50 years old, after 50 years old both women and men start to develop DS, with women having a faster developing rate than men. For elderly Chinese (≥ 65 yrs, mean age: 72.5 yrs), large population based studies (MsOS(Hong Kong) and MrOS(Hong Kong), females n=2000 and males n=2000) showed DS prevalence was for 25.0% for women and 19.1% for men, and the prevalence F:M (women:men) ratio was 1.3:1. The published data (MsOS(USA) and MrOS(USA) studies) seem to show elderly Caucasian American has a higher DS prevalence, being approximately 60-70% higher than elderly Chinese; however the prevalence F:M ratio was similar to elderly Chinese population. Patient data showed female patients more often received treatment than men; and preliminary data show the ratio of numbers of female patients received treatment compared with men did not differ between Northeast Asians (Chinese, Japanese, and Korean) and European and American Caucasians, being around 1.9:1 in elderly population. Compared with Caucasians, Asians are likely to have Meyerding II slip and multi-level slips, and receive treatment more than half decade earlier. The existing data also suggest that menopause may be a contributing factor for the accelerated DS development in post-menopausal women.

## 1. Introduction and basic concepts

Degenerative spondylolisthesis (DS) is a disorder that causes the slip of one vertebral body over the one below due to degenerative changes. It differs from spondylolytic spondylolisthesis by the absence of a pars interarticularis defect (spondylolysis), i.e., in DS the whole upper vertebra (vertebral body and posterior part of the vertebra including neural arch and processes) slips relative to the lower vertebra. Both DS and spondylolytic spondylolisthesis are commonly seen as incidental findings in asymptomatic patients. An understanding of the natural history of these conditions is important to counsel patients and determine a course of actions. The plain radiographic features include the essential finding of spondylolisthesis on a lateral view of forward (or backward) displacement of L4 on L5 or, less commonly, L5 on S1 or L3 on L4 in the presence of an intact neural arch. The listhesis is a rotary deformity and not a simple forward (or back forward) displacement [1]. Radiograph can show small compensating curves in the upper lumbar and lower thoracic spine [1]. The major local reasons of DS that probably lead to the development of degenerative vertebral slippage are: (1) arthritis of the facet joints with loss of their normal structural support; (2) malfunction of the ligamentous stabilizing component, probably due to hyperlaxity; and (3) ineffectual muscular stabilization [2-7]. Disc degeneration leads to segmental instability in the sagittal plane and may result in DS [8]. Pregnancy and sports activities are also associated with DS [9-14].

Separation of the pars interarticularis can occur when spondylolysis is present (Fig 1). Spondylolysis can be congenital or caused by a stress fracture of the bone, and is especially common in adolescents who over-train in activities [2, 12-14]. The pars interarticularis is vulnerable to fracture during spinal hyperextension, especially when combined with rotation, or when experiencing a force during a landing. This stress fracture most commonly occurs where the concave lumbar spine transitions to the convex sacrum (L5-S1). A significant amount of individuals with spondylolysis will develop spondylolisthesis, counting for 50-81% of this particular population. It is believed that both repetitive trauma and an inherent genetic weakness can make an individual more susceptible to spondylolysis [15-16].

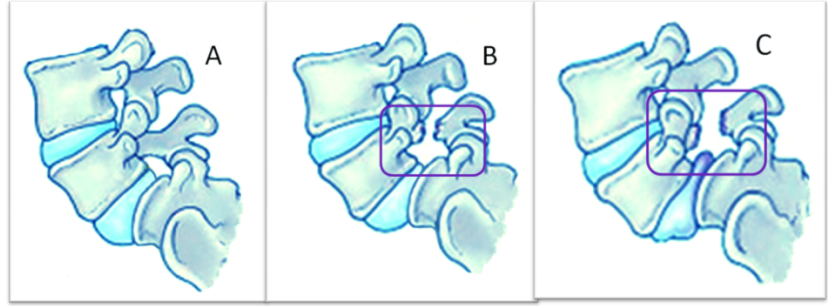

Fig 1, A: normal anatomy L5/S1, B & C: different extent of spondylolytic spondylolisthesis of L5/S1.

After degenerative changes unlocked the intervertebral joint, the vertebral body slipping occurs along a direction that roughly depends on two factors: (1) the symmetry of facet joint lesions and (2) the distribution of weight bearing forces. When facet joint subluxation is symmetric, slipping is mainly sagittal; while with asymmetric subluxation, a rotatory displacement also occurs. Defect of pars interarticularis seen on lateral or bilateral oblique views helps to distinguish between DS and isthmic spondylolisthesis. The additional findings include disc space narrowing, endplate sclerosis, peridiscal osteophytes, and facet sclerosis and hypertrophy. In the last stage, osteophytes and advanced disk space narrowing lead to restabilization of the intervertebral level with decrease or disappearance of the range of movement [17, 18].

The natural course of spondylolysis and spondylolisthesis has been studied [19-21]. A prospective study was initiated in 1955 with a radiographic and clinical study of 500 first-grade children. A total of 22 subjects 6 years of age were determined to have a lytic defect of the pars interarticularis, giving a rate of 4.4%. Till adulthood, 30 individuals were identified to have pars lesions with a prevalence rate of 6%, consisted of 10 females and 20 males (females:males ratio, F:M ratio 1:2). Of the 30 subjects, 22 had bilateral L5 pars defects and 8 subjects had unilateral defects. All bilateral pars defects were at L5. Over the course of the study, spondylolisthesis developed in 18 of the 22 subjects with bilateral L5 lesions (spondylolisthesis prevalence = 18/500 =3.6%). The average slip was 11% for all subjects with initial spondylolisthesis. The average slip in the 1999 studies for this group was 18%. There appears to be a marked slowing of slip progression with each decade, while no subject has reached a 40% slip. It was suggested most patients live active, pain-free, fully functional lives. There is no correlation between changes in clinical symptoms and progression of spondylolisthesis. DS, although progressive, rarely exceeds grade II. Low-grade isthmic spondylolisthesis rarely progresses, and it has a benign clinical course in the majority of patients [20]. Spur formation, sclerosis, and ossification of ligaments limit the progression of degenerative spondylolisthesis. Fredrickson *et al* [21] suggested that a child with spondylolysis or spondylolisthesis can be permitted to enjoy a normal childhood and adolescence without restriction of activities and without fear of progressive olisthesis or disabling pain.

Progressive change in the intervertebral disc, thickening of the ligamentum flavum, and translation of the vertebra all contribute to the compromise of the canal and central spinal stenosis. The process may also cause foraminal narrowing due to the impingement of the superior articular process in the neuroforamina. Patients presenting with DS may have any combination of low back pain, neurogenic claudication, and radiculopathy [22]. Hamstring spasm is the most frequently associated neurologic abnormality. Lumbar radiculopathy and bowel or bladder symptoms are rare but may occur in individuals with severe isthmic spondylolisthesis. In the case of DS with nerve root compression, the anatomic type of vertebral slippage has an influence on the pattern of the neurologic symptoms. In the case of symmetric spondylolisthesis, with no or only mild rotatory component, nerve root involvement that is typically bilateral and pluriradicular is related to the compression of the thecal sac in the central spinal canal and to a bilateral lateral recess stenosis. Conversely, in the case of asymmetric spondylolisthesis, with a marked rotatory displacement, nerve root involvement is frequently unilateral involving one or two nerve roots on the side of maximal facet joint subluxation because of the compression of the nerve roots in the ipsilateral lateral recess and foramen. In a meta-analysis dealing with surgically treated DS, radiculopathy and neurogenic claudication were found preoperatively in 32% and 3% of the patients, respectively [23]. Harris and Weinstein [24] studied the long-term outcome in patients with Meyerding grades III or IV spondylolisthesis (≥51% slip) and found that 36% of the patients treated nonsurgically were asymptomatic, 55% had occasional back pain, and 45% had neurologic symptoms; none of the patients was incontinent.

Only 10–15% of patients seeking treatment eventually will have surgery [25]. Also, a multilevel slip, or a greater degree of slip (> 25%) do not increase the prevalence and/or severity of symptoms [26, 27]. In symptomatic patients with DS whose debilitating condition is nonresponsive to conservative management, surgical intervention is performed. For the clinical management of spondylolysis and spondylolisthesis, readers are advised to refer to references [28-30].

2. Classifications

The Wiltse-Newman classification (Table 1) is the most widely used classification of spondylolisthesis [31]. Of the five types, types I and II apply commonly to the child and adolescent. Type I, the dysplastic type, defines spondylolisthesis secondary to congenital abnormalities of the lumbosacral articulation, including maloriented or hypoplastic facets and sacral deficiency. The pars is poorly developed, which allows for elongation or eventual separation and forward slippage of L5 on the sacrum with repetitive loading over time. Type I is less common, comprising 14% to 21% of congenital cases [32, 33]. Type II, the isthmic type, defines spondylolisthesis that results from defects of the pars interarticularis. This group is subdivided into three subtypes. Type IIA, the most common subtype, is caused by fatigue failure of the pars from repetitive loading, resulting in a complete radiolucent defect. Type IIB is caused by an elongated pars secondary to repeated microfractures that heal. This type can be difficult to distinguish radio graphically from the dysplastic type. Type IIC refers to a pars fracture that results from an acute injury. Wiltse hypothesized that isthmic defects are the result of chronic loading of a pars interarticularis that is genetically predisposed to fatigue failure [34]. Marchetti and Bartolozzi [35] proposed an alternative classification system with two broad categories developmental and acquired. The developmental category defines spondylolisthesis resulting from an inherited dysplasia of the pars, lumbar facets, discs, and vertebral endplates, combining the dysplastic and isthmic categories of Wiltse-Newman. Acquired spondylolysis and spondylolisthesis define failure of the pars secondary to repetitive spinal loading related to specific activities.

Table 1: Classification system for spondylolisthesis

| Wiltse-Newman | Marchetti-Bartolozzi |
|---|---|
| I. Dysplastic | Developmental |
| II. Isthmic |     High dysplastic |
|     IIA, Disruption of pars as a result of stress fracture |         With lysis |
| |         With elongation |
| |     Low dysplastic |
|     IIB, Elongation of pars without disruption related to repeated, healed microfractures |         With lysis |
| |         With elongation |
| | Acquired |
|     IIC, Acute fracture through pars |     Traumatic |
| III. Degenerative |         Acute fracture |
| |         Stress fracture |
| IV. Traumatic |     Postsurgery |
| V. Pathologic |         Direct surgery |
| |         Indirect surgery |
| |     Pathologic |
| |         Local pathology |
| |         Systemic pathology |
| |     Degenerative |
| |         Primary |
| |         Secondary |

In dysplastic spondylolisthesis (Wiltse-Newman type I), the L5 vertebra with intact posterior elements slips forward on the sacrum. The resulting lumbar stenosis may cause L5 nerve radiculopathy as well as bowel and bladder dysfunction from compression of sacral nerve roots. Children and adolescents with dysplastic spondylolisthesis are more like to develop neurologic injury and carry greater risk of progressive deformity than do patients with isthmic spondylolisthesis (Wiltse-Newman type II). McPhee *et al* [36] reported a higher frequency of progression in the dysplastic type (32%) than in the isthmic type (4%). Furthermore, patients with dysplastic spondylolisthesis are more likely to require surgical treatment [32, 37]. Children who are diagnosed before their adolescent growth spurt, girls, and those presenting with >50% slip are most likely to progress [38]. Spondylolisthesis associated with congenital dysplasia of the lumbosacral facets and sacrum allows anterior translation of the L5 vertebral body with intact posterior elements that can compress the L5 and sacral nerve roots.

3. **Radiodiagnostics**

Radiograph is the first line investigation for suspected spondylolisthesis (Fig 2). The best approach to diagnose spondylolisthesis remains controversial, and it also remains unknown whether the selection of radiodiagnostic techniques will influence clinical management. Standing postero-anterior and lateral radiographs of the thoracolumbar spine, with supine oblique views of the lumbosacral spine, are usually used to assess potential spondylolysis or spondylolisthesis. The standard posteroanterior radiographic view allows evaluation of coexisting scoliosis that may be secondary to paraspinal spasm, whether idiopathic or olisthetic (i.e., the result of asymmetric forward vertebral translation at the level of the spondylolisthesis). The standing lateral view is useful for identifying spondylolytic defects and documenting the degree of spondylolisthesis. The abnormal translation may increase in standing position compared to recumbent position. Vertebral sagittal slip is higher in standing position than in recumbent supine position [39]. In the coronal plane, a slight disruption of the alignment of the spinous processes and of the lateral border of the vertebral bodies with or without a lateral slip (laterolisthesis) should be checked carefully. Supine oblique and spot lateral radiographic views of the lumbosacral junction improve the likelihood of diagnosing stress reactions and spondylolytic defects. Harvey *et al* [40] suggested the coned lateral of the lumbosacral junction and anterioposterior (AP) view with 30 degree cranial angulation are particularly important. Plain radiography of the pars is sometimes difficult since the pars lies oblique to all three orthogonal planes. A trapezoidal shape of the fifth lumbar vertebra is probably a result of the slipping, not a cause [21]. Spina bifida occulta occurs more frequently in patients with a pars interarticularis defect than in patients without a defect.

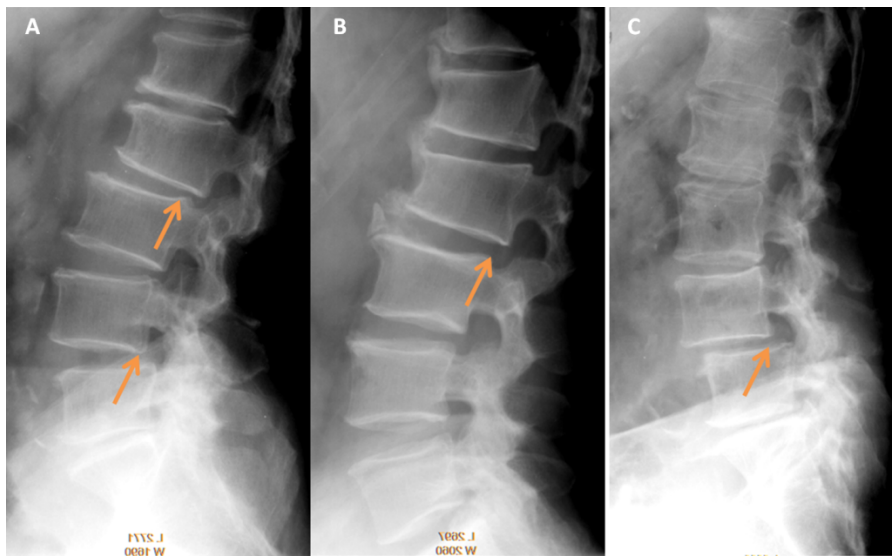

Fig 2. A: multiple-level spondylolisthesis of L2 (Grade-I posterolithesis) and L4 (Grade-I anteriolithesis); B: spondylolisthesis of L2 (Grade-I posterolithesis), with formation of osteophytes probably as a mechanism to compensate for stabilization; C: L4 spondylolisthesis (Grade-I anteriolithesis).

Flexion-extension positioning is a technique used by many surgeons to assess the degree of lumbar instability in spondylolisthesis [41]. However, it has been speculated that tightness or spasm in the paraspinal muscles secondary to pain when standing may have a splinting effect, thus reducing the apparent instability on flexion-extension radiographs. Pain leads to decreased intervertebral motion in symptomatic patients with spondylolisthesis [42]. While static radiographs seem to show the greatest slip in standing position, in cases of hypermobile spondylolisthesis the lateral decubitus position may reveal an even higher abnormal translation, because the stiffness of the splinting muscles is reduced [43]. Adequate functional radiographs depend on the patient's cooperation, on the examiner's proper control, and can lead to different results from test to test [44]. According to Danielson *et al* [45,46], a slight variation in patient positioning or in gantry tilting may result in a 10% to 15% variation in the range of vertebral displacement. Patient positioning and direction of the X-ray beam have to be accurate and reproducible to allow optimal measurement. However, the way to perform functional radiographs and the method to measure displacements are still not standardized. Anderson *et al* [47] reported initial radiographs and nuclear planar bone scans failed to demonstrate 19% of the pars lesion in one their study.

Factors required additional radiological investigations include significant and progressing neurologic claudication or radiculopathies and clinical suspicion of another conditions, such as metastatic disease, may be causative. An absolute indication is the presence of bladder or bowel complaints [48]. Additional studies that may be selected include technetium bone scanning, particularly when a metastatic tumor is suspected. Single-photon emission CT (SPECT) of the lumbosacral spine is the most effective method for detecting spondylolysis when plain radiographs are normal and the patient history and physical examination are suggestive of the diagnosis. Increased radionuclide uptake in an intact pars, lamina, or pedicle is consistent with a stress reaction. A relative decrease in tracer uptake on serial SPECT scans has been correlated with improvement of clinical symptoms and signs in patients treated for symptomatic [49]. As spondylolisthesis develops and progresses, the SPECT scan again becomes positive.

SPECT scanning in spondylolysis is not a positive or negative process, but rather varies with the time and stability of the spondylolytic spine.

Stress reactions that have not progressed to complete defects will be radiographically occult. If the radiographic series is nondiagnostic, limited thin section CT technique will demonstrate both stress reactions and established defects, though the diagnosis of spondylolysis being the cause of pain is more difficult. Thin-section CT, performed with a reverse gantry angle, is the best modality for defining the bony anatomy of spondylolysis [50]. Magnetic resonance imaging (MRI) is indicated when neurologic symptoms and signs are present in conjunction with spondylolysis and spondylolisthesis. Nerve root compression, lumbar disk abnormalities, spinal cord anomalies, and neoplasm of the spinal cord or vertebral spinal column are other sources of low back pain that are best assessed with MRI [51]. MRI may demonstrate intraosseous edema in the affected areas in these patients.

A few methods have been proposed to grade spondylolisthesis. The first is the method of Meyerding [Fig 1, 52]. The antero-posterior (AP) diameter of the superior surface of the lower vertebral body is divided into quarters and a grade of I–IV is assigned to slips of one, two, three or four quarters of the superior vertebra, respectively [Fig 2]. The second method, first described by Taillard [53], expresses the degree of slip as a percentage of the AP diameter of the top of the lower vertebra. The second method is favored by most authors as it is more accurately reproducible [48]. A simpler classification system divides spondylolisthesis into cases with translation of ≤50% (stable) and those with translation of >50% (unstable). Patients with higher grades of spondylolisthesis and higher slip angles, a measure of lumbosacral kyphosis, have a higher risk of progression. Measurement of the slip and its apparent progression, however, should be viewed with caution. Studies have shown that there can be inter and intra-observer error of up to 15% [45, 46]. This variation can increase if there is an element of rotation. The position of radiograph at different time points will also affect the measurement results.

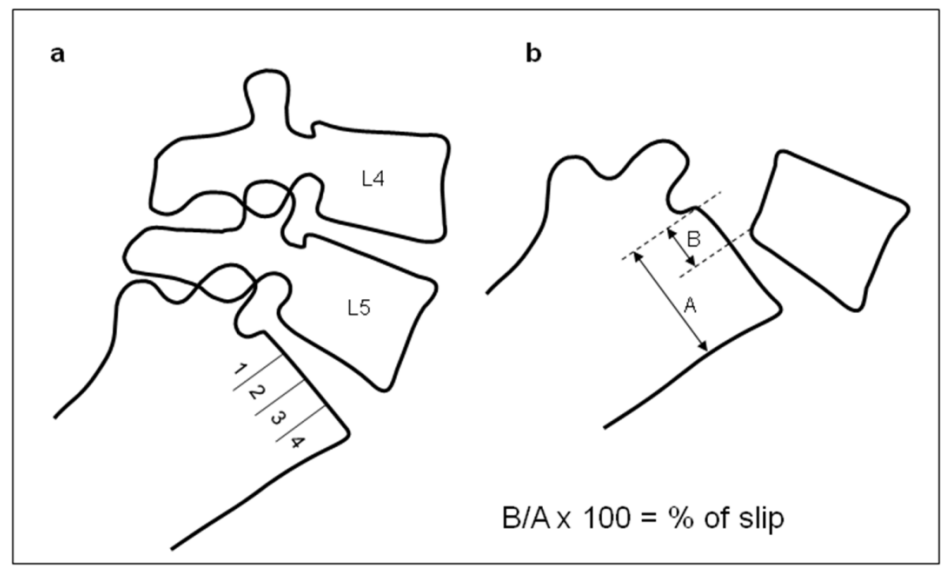

Fig 3, Scheme of spondylolisthesis grading methods, a: Meyerding, b: Taillard.

For more discussion on radiological investigations of spondylolisthesis, see references [50, 51, 54-56].

4. **Epidemiology I: results of population based studies.**

The published epidemiological data on DS varies greatly one from another. The prevalence, female male prevalence ratio, as well as risk factors remain controversial. For example, in an elderly Chinese population (≥65 yrs) of 4000 subjects (half females and half males, mean age: 72.5 years), it was found the overall prevalence of spondylolisthesis was 25.0 % for females and 19.1 % for males, with a F:M ratio of 1.3:1 [27]. In the Copenhagen Osteoarthritis Study (1533 males, mean age of 62 years, range 23–93 yrs; 2618 females, mean age: 65 years, range 22–9yrs), Jacobsen *et al* [57] reported the prevalence of spondylolisthesis was 2.7% for males, and 8.4% for females, with F:M ratio of 6.4:1. Farfan studied 460 lumbar spines autopsy (mean age: 64 yrs), found the prevalence of DS was 4.1% [1]. In a professional taxi driver cohort in Taipei (mean age: 44.5 ±8.7 yrs, predominately males), Chen *et al* [58] reported the prevalence of spondylolisthesis was 3.2%. Kalichman *et al* [59] studied 188 adults community-based population (mean: 52.7 ±10.8 years) with CT, the DS prevalence was 7.7% (males) vs. 21.3% (females, F:M ratio=3:1). This controversy is complicated by the facts that imaging techniques and radiographic landmarks used for the measurements vary between reported studies and radiographic magnification is not always taken into account. For this review we searched systemically the published literatures, with the aim to 1) have a clearer understanding of age specific and gender specific prevalence of DS; 2) whether there is a prevalence difference of DS between Northeast Asians (Chinese, Japanese, and Korean) and European/American Caucasians; and 3) the potential of menopause for DS development.

PubMed was used to search literature. To broadly include data only the word 'spondylolisthesis' was used for search. The search was carried out on April 24, 2016, and there were 5255 items found. This was again updated on September 18, 2016, with 120 new items found. These total 5375 items was evaluated by the authors according their relevance to the current study. Relevant articles containing original data were reviewed. The emphasis of this study in degenerative spondylolisthesis, therefore papers solely deal with congenital spondylolisthesis was not selected. Spondylolysis without listhesis, iatrogenic spondylolisthesis [60], traumatic spondylolisthesis, and spondylolisthesis among athletes were not the focus of this literature survey.

Reported data showed the fetal incidence of spondylolisthesis (i.e. with a vertebra slip) is close to zero [21, 61-63]. The incidence of dysplastic or isthmic spondylolisthesis in the general population is 4-8% [19, 59, 61-67]. Beutler *et al* [19] reported that lytic defect of the pars interarticularis prevalence rate was 6%, and the spondylolisthesis prevalence was 3.6%. The male prevalence is likely to be twice of that for female [19]. In a study involving 2000 Japanese general population with multidetector computed tomography scan, Sakai *et al* [66] reported lumbar spondylolysis was found in 5.9% subjects, and F:M ratio was 1:2. They also noted spondylolisthesis was found in 74.5% of the subjects with bilateral spondylolysis, and in 7.7% of those with unilateral spondylolysis. This study suggests shows Caucasian and Japanese may have similar congenital spondylolysis prevalence. One study reported CT show higher spondylolysis rate (11%) than radiograph [59], however this was not confirmed in other CT based studies [66, 67]. Some ethnics groups may have higher prevalence rate and that may be related to generic factors [68-71], such as a higher prevalence in Inuit population and in black American female population has been reported [64, 68,72].

Our review showed DS is strongly age specific and gender specific, and it is relatively rare before 50 yrs old (Fig 4). In the Copenhagen Osteoarthritis study DS increases with increased age in both sexes, while very few individuals (about 4% of all DS cases) had spondylolisthesis at L4 to L5 level before age 50 (Fig 4a). Kalichman *et al* [59] also reported no cases of DS were observed in men less than 40 years or in women less than 50 years of age (Fig 4b). Similar result was shown in Chen *et al*'s study [58]. The degree of DS slip is usually grade-I [26, 27]. The DS level most commonly involved was L4–L5, followed by L5–S1 and L3–L4 (approximately 12% each) [26, 27, 73]. It has been demonstrated that the progression of slipping is slow and not correlated to age at diagnosis and initial degree of spondylolisthesis. Disc height reduction at the spondylolytic level occurred at an earlier age and was more severe than in a normal group. Symptoms were correlated to radiographic pathology. Risk factors for low-back symptoms were greater than 25% slipping, spondylolysis at the L4 level, and early disc degeneration [75-77].

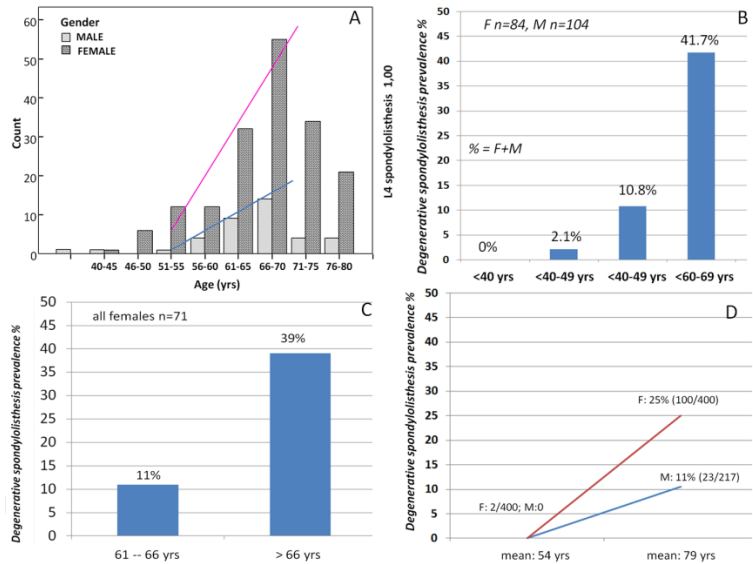

Fig 4, The prevalence of DS is very gender-specific and age-specific. Both women and men have few DS before 50 years old. After 50 years old both women and men start to develop DS, with women having faster developing rate than men (A, D). A: raw data from reference 57, B: raw data from reference 59, C: raw data from reference 78, D: raw data from reference 26.

With population based epidemiology studies of MsOS(Hong Kong) and MrOs(Hong Kong) [27], 2000 elderly Chinese men and 2000 elderly Chinese women were recruited during the period of August 2001 to March 2003 [27]. The data showed the prevalence of lumbar spondylolisthesis was shown to be 25.0% in women, and 19.1% in men, and the F:M was ratio of 1.3:1 (table 2). These studies did not differentiate between DS and spondylolytic spondylolisthesis as only lateral radiography was obtained, and therefore our data probably overestimate prevalence by 4% for this reason. It is also probably that these data underestimate the prevalence by around 8% because standing radiograph was not obtained [37, 41, 53]. However, this limitation is expected would not affect the F:M ratio. Hensinger [37] reported that the average slip of 9% on the supine radiographs increased to 10% on the erect views and this was not significant, while in a low back pain patient cohort Iguchi *et al* [41] reported that 15% of spondylolisthetic lesions can only been seen by a standing lateral. MsOS(Hong Kong) and MrOs(Hong Kong) data could not be compared with Chen *et al*'s results as their subjects were younger [58]. With a small cohort (male n = 306, female n = 486) population based study reported by Chaiwanichsiri *et al* [79], the prevalence was 14.4% (females age: 60.88±7.9 yrs) vs. 8.8% (males 61.39±7.7yrs), with F:M ratio of 1.63:1. In MsOS(Hong Kong) and MrOs(Hong Kong) cohort aged 65-69 yrs, prevalence was 21.1% (females) vs. 14.7% (males), with F:M ratio of 1.45:1 [27]. These data are probably comparable since our subjects were older than Chaiwanichsiri *et al*'s.

The design of MsOS(Hong Kong) and MrOS(Hong Kong) studies is similar to MsOS(USA) and MrOS(USA) studies [27, 73, 80, 81]. In the studies spondylolisthesis was defined as a forward slip (anterolisthesis) or backward slip (retrolisthesis) of one vertebral body by at least 5 % in relation to the next most caudal vertebral body, this would roughly translated to 3mm. For women, in the MsOS (USA)study (n=788 subjects analyzed) study Vogt *et al*. [79] used greater than 3 mm as the threshold of spondylolisthesis and measured at the lower lumbar level (L3 to S1). The prevalence of DS was 43.1% (anterolisthesis=28.9 %, retrolisthesis = 14.2%), higher than Hong Kong results of 25% (table 2). Compared with the MrOS (USA) study (n=295 subjects analyzed), Hong Kong male cohort also had a lower anterolisthesis prevalence (19.1 % versus 31 %) [27, 81]. These results suggest elderly Caucasian American has higher DS prevalence, approximately 60-70% higher, than elderly Chinese. However, the F:M ratio was similar to Chinese population, being 1.38:1 (table 2). The 4-years *de novo* DS is also similar between Chinese men and Caucasian American men, being around 12%. That the Caucasian has higher DS prevalence has been indicated, in smaller cohort studies, by Kalichman *et al* [59, Fig 4b] and Marty-Poumarat *et al* [78, Fig 4c]. On the other hand, Japanese population may also have a lower DS prevalence than Caucasian. For the elderly population from a single Japanese village with 205 elderly men (mean age, 70.7 years) and 323 elderly women (mean age, 70.5 years), in a cross-sectional study Horikawa *et al* [82] reported spondylolisthesis prevalence of 4.9% for males, and 11.5% for females. In a cohort of 3,259 Japanese patients with low back and/or leg pain (mean age approximately 65 years), Iguchi *et al* [41] reported a DS prevalence of about 8.7% (F:M ratio =1.3:1). Additionally, Aono *et al* [83] followed up for 12.1 years (8–14 years)

17in 142 female subjects without spondylolisthesis at baseline radiographs (mean baseline age: 54.7 years, range: 40–77yrs), and the incidence of newly developed DS was 12.7%. The same group (Kobayash *et al*, 84) reported when the female subjects was 68.5±9.2 years old, the DS prevalence was 24.8% (50 subjects out of 289). These data suggest that elderly Japanese females and elderly Chinese females in Hong Kong have similar DS prevalence and progression rate. Of note, a few studies showed body height was not associated the incidence of DS [27, 57, 73]. Combining baseline and yr-4 follow-up data, the elderly Chinese DS prevalence based on elderly Chinese MsOS(Hong Kong) and MrOS(Hong Kong) studies is shown in Fig 5.

Table 2. A comparison of degenerative spondylolisthesis prevalence and 4-years progression in elderly Chinese and elderly Caucasian American [27, 73, 80, 81].

|  | Age (yrs) | prevalence | progression | *de novo* |
|---|---|---|---|---|
| Ms OS (Hong Kong) yr-0 [a] | 72.6 (range 65–98) | 25% |  |  |
| MsOS (USA) yr-0 [c] | 71.5 (range: 65-89) | 43.1% |  |  |
| MrOS (Hong Kong) yr-0 [b] | 72.4 (range 65–92) | 19.1% |  |  |
| MrOS (USA) yr-0 [d] |  | 31% |  |  |
| MsOS (Hong Kong) yr-4 [e] | 75.7 (range:68–102) | 41.2% | 16.5% | 12.7% |
| MrOS (Hong Kong) yr-4 [f] | 75.5 (range:68–95) | 31.5% | 13.0% | 12.4% |
| MrOS (USA) yr-4 [g] |  | 43%[h] | 12% | 12% |

[a]n=1994 subjects, [b] n=1996 subjects, [c]n=788 subjects, [d]n= 295 subjects, [e]n= 1546 subjects, [f]n= 1519 subjects, [g] n=190 subjects, [h]: estimated from baseline data plus *de novo* number. The F:M ratio of MsOS(Hong Kong) and MrOs(Hong Kong ) is 1.3:1; while the F:M ratio of MsOS(USA) and MrOs(USA) is 1.38:1.

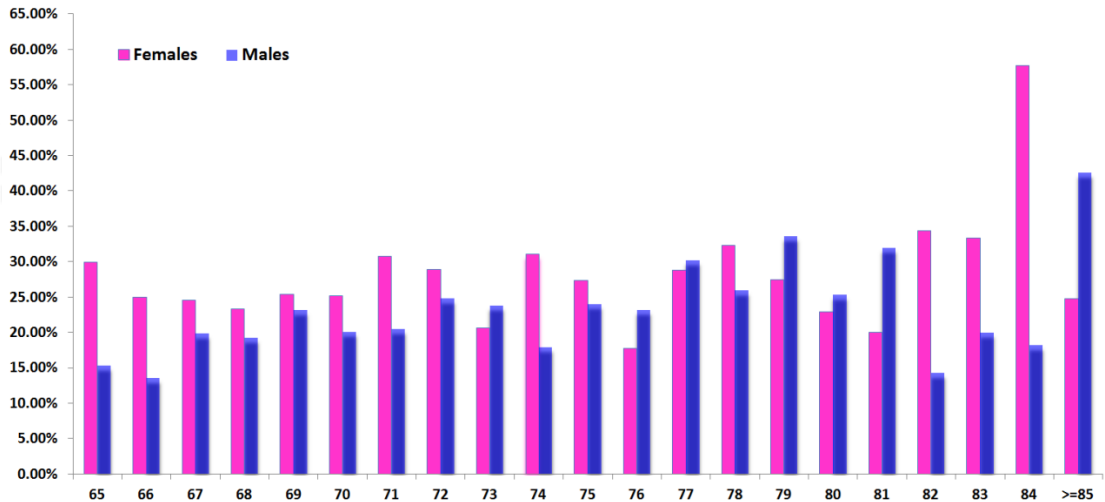

Fig 5, The elderly Chinese DS prevalence based on elderly Chinese MsOS(Hong Kong) and MrOS(Hong Kong) studies. Baseline data and yr-4 follow-up data have been combined. X-axis: age in yrs; Y-axis: prevalence (in %). Source data from references 27 and 73.

Based on the observations above (table 2, Fig 4), we can conclude that elderly Chinese, probably all Northeast Asians, has a substantial lower prevalence of DS than European and American Caucasians. The precise difference is difficult to confirm due to the factor that the DS prevalence is very age-specific and gender-specific, and few age-matched studies are available for comparison. The results of Table 1 suggest that F:M ratio (=1.38:1) in elderly Caucasian American is similar to Chinese, in contrast to the much larger F:M ratios reported by some group [57, 64]. Iguchi *et al*'s [41] low back and/or leg pain cohort also had a F:M ratio of 1.3:1 (115 women and 86 men). Of note, Meana *et al*'s study on Canadian women reported that in the age group 65 years and older Chinese males and females had the lowest rates of chronic pain than other ethnic groups [85]. Deyo *et al* [86] also estimated back pain prevalence and visit rates from U.S. national surveys 2002, and showed among racial groups, American Indians and Alaska natives had the highest prevalence of back pain, while Asian Americans had the lowest prevalence (Asian Americans vs. non-Hispanic Whites=19.0% vs. 27.2%, or 1: 1.43).

### 5. Epidemiology II: ratio for female and male patients received treatment

Previous study showed elderly Chinese, Japanese and Korean women have very similar age-specific osteoporotic vertebral fracture prevalence, but elderly Caucasian women have slight higher age-specific osteoporotic vertebral fracture prevalence elderly than Northeast Asians [the F:M ratio for elderly American females and elderly Chinese females is around 1.3:1, table 6 of reference 87]. Patient series on European Caucasian or American (the majority being whites) or Northeast Asian (Chinese, Japanese, and Korean) were retrieved for the current study. Including criterion for this literature survey was gender specific age information was available in the publications, or the age distribution was rather narrow, so that female patients and male patients were close to be age-matched and F:M ratio can be computed.  We only included the series which have generally no less than 24 patients. The result is shown in Figure 6.

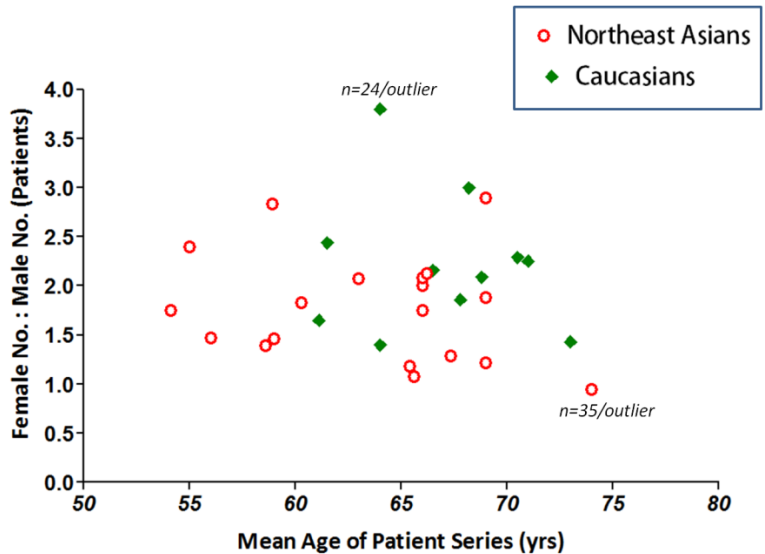

Fig 6, Patient series age-specific female:male ratio separated into European/American Caucasian and Northeast Asian groups. Data were extracted from references 88-114. There were 19 patient series for Asians and 12 patient series for Caucasians. Excluding two outliers, the mean age were 63.0±5.10 yrs for Asians and 67.6±2.58 yrs for Caucasians respectively (p=0.01); and the mean F:M ratio were 1.81±0.53 for Asians and 2.0±0.34 for Caucasians respectively (p=0.22).

The Figure 5 data show the patients were mostly > 50 years old, and Asian patients were treated earlier than Caucasians. It is unknown whether Asians were likely to be symptomatic earlier, or they had easier access to early treatment. However, compared with the MrOS(USA) study, MrOS(Hong Kong) cohort had more Meyerding grade II anterolisthesis (2.8% vs. 1%) and slightly more subjects with more than one level of anterolisthesis (5.1% vs. 4%) [27, 81]. For the females, 5.2 % of MsOS(Hong Kong) cohort had Meyerding grade II slip, while only 1% of MsOS(USA) had Meyerding grade II slip. 13.8 % of MsOS(Hong Kong) cohort had slips at two or more levels, this figure was close to 10% for MsOS(USA) cohort [27, 80]. The data presented by Iguchi *et al* [41] also seem to concur with MsOS(Hong Kong) and MrOS(Hong Kong) studies. Therefore, Asians were likely to be symptomatic earlier than Caucasians.

That higher F:M ratio shown in Figure 5, approximately 2:1, is than population based ratio (1.3:1). This can be explained that women has lower threshold value for back pain, and women are more likely to seek pain treatment than men [115]. There is a higher prevalence of pain in females for headache, migraine, temporomandibular pain, burning mouth pain, neck pain, shoulder pain, back pain, knee pain, abdominal pain, and fibromyalgia [115]. Women have shown to have a lower threshold of perception of pain and in reaction to it [116-19]. It has been observed that although females are more likely to report symptoms, even when physician verified abnormalities are approximately equal to those of males [120-121]. There are more female patients than male patients in general pain clinics [122, 123].

There are a number of the limitations of this patient series review. The majority of the published papers did not separately report the age information for females and males. The age of females and males of some publications might be actually marched even without the exact necessary information provided; however in these cases we still could not use these papers for Figure 5. Also we founds the majority of patient series reported in Chinese language did not separate isthmic spondylolisthesis and degenerative spondylolisthesis. Some of the patient series were limited in patient number and the F:M ratio might be coincidental. In some series the F:M ratio might be due to the type of treatment selected. We searched items in English, Chinese, German, French, Spanish, and Hungarian; publications in other some languages such as Polish were not analyzed. More studies are required to further confirm the pattern seen in Figure 5.

## 6. The role of menopause on degenerative spondylolisthesis

The association between menopause and DS was considered due to the higher incidence of DS in post-menopausal women than in age-match men. Before the age of fifty years, DS is rare; and the prevalence of congenital spondylolisthesis is actually more common in men as discussed above [19, 66]. Low level of female sex hormones in post-menopausal women can be associated with 1) accelerated degeneration of disc degeneration and disc space narrowing [124-128]; 2) higher prevalence of osteoarthritis, including that of facet joints [129,130]; 3) general laxity of the paraspinal ligaments. It has been shown that HRT preserves muscle strength in postmenopausal women [131, 132]. Taaffe *et al* [133] showed that hormone replacement treatment (HRT) preserves or improves skeletal muscle quality in early postmenopausal women, and has a positive effect on muscle performance. In experiments on the elastic properties of the capsular ligament of the hip, periodontal tissues and the uterus, oestrogen was proved to have a considerable influence on collagen and elastin synthesis [134-137]. Fischer and Swain [138] found that oestrogen reduced the collagen content and increased the elastin content of the arterial wall in rats. Shikata *et al* [134] in an experimental study on hip dislocation stated that oestrogen deficit after oophorectomy induced a loss of elasticity of the capsular ligament.

In 1994 Imada *et al* [139] performed a case-control study on the influence of oophorectomy on the development of DS. The mean period between oophorectomy and review was 6.3±2.8 years. They found bilateral oophorectomy with no hormonal replacement therapy was a risk factor for DS with an odds ratio of 7.5 (95% confidence interval, 1.6 to 46). The incidence (n=20) of DS in 69 oophorectomised patients (mean age at oophorectomy: 47.4±5.6 yrs) was about three times higher than in 69 non-oophorectomised matched control subjects (n=6, 29.0% vs. 8.7%, p < 0.005). Their results suggest that the abrupt decrease in oestradiol level caused by oophorectomy may be a predisposing factor in degenerative spondylolisthesis at L4/5. The loss of elasticity in the paraspinal ligamentous system produced by the hormonal changes caused by oophorectomy may contribute to degeneration and to the development of the vertebral slip at L4/S. Of note, bilateral bilateral oophorectomy has acute and dramatic impact on the physiology of musculoskeletal system [140, 141].

Marty-Poumarat *et al* [78] evaluated the influence of hormone replacement therapy (HRT) on degenerative scoliosis and on lateral rotatory olisthesis (LRO). A cross-sectional study was conducted in 146 postmenopausal women: 75 women had received HRT for more than 1 year (HRT>1, age: 65.0±5.4 yrs) and 71 women had never received HRT or less than 1 year (HRT<1, age: 66.7±5.9 yrs, p=0.07). All the women had been menopaused for more than 5 years. There was no difference in body mass index in the two groups (24.2± 3.4 vs 25.0±2.9 kg/m2, p=0.11), and the mean time since menopause was not significantly different in the two groups (14.8±6.3 vs.16.8±7.2 0.09 yrs, p =0.09). The HRT duration was 8.7 ± 6.1 yrs for HRT>1 group. The prevalence of LRO was significantly lower in HRT>1 group than in HRT<1 group (8% vs. 30%). LRO increased with age only in HRT <1 group (11% when aged ≤66 years vs. 39% when aged >66 years, p = 0.013), whereas the prevalence of LRO remained stable in HRT>1 group.

Our MsOS(Hong Kong) and MrOs(Hong Kong) studies demonstrated the modifiable factor for DS occurrence include to control the body weight and have moderate exercises to increase protective muscle strength, and also control blood pressure, particularly for women [27, 73]. HRT initiated at early postmenopausal phase may be protective for intervertebral disc degeneration [142], osteoarthritis of facet joints, and maintain muscle and ligament tone in lumbar regions, as well as attenuate atherosclerosis [143]. HRT regimen may be considered in cases of women with anatomical high risks of developing DS or DS progression, such as the factors of high lumbar lordosis, vertebral end-plate inclination, severe disc degeneration and loss of height, and facet joint sagittal orientation [3, 7], particularly for cases with clinical symptoms. Optimal selection of dose regimen, combination of oestrogen with progestins versus oestrogen alone, the administration route, and duration of treatment such as the choice of repetitive or periodic administration simulating the menstrual cycle may lead to better benefits [144, 145]. Women presenting to their medical providers during the menopausal transition provide a unique opportunity for risk assessment, counseling and the institution of various prevention measures [145].

In conclusion, our review demonstrated the prevalence of DS is very gender specific and age specific. Both women and men have few DS before 50 years old, after 50 years old both women and men start to develop DS, with women having faster developing rate than men. Elderly Caucasian American has a higher DS prevalence, being approximately 60-70% higher, than elderly Chinese. The women:men prevalence ratio is estimated to around 1.3:1. Further studies on DS epidemiology should report gender specific and age specific information to allow better inter-study comparison and data synthesis. Evidences suggest that hormone replacement therapy may alleviate the development of DS in postmenopausal women.

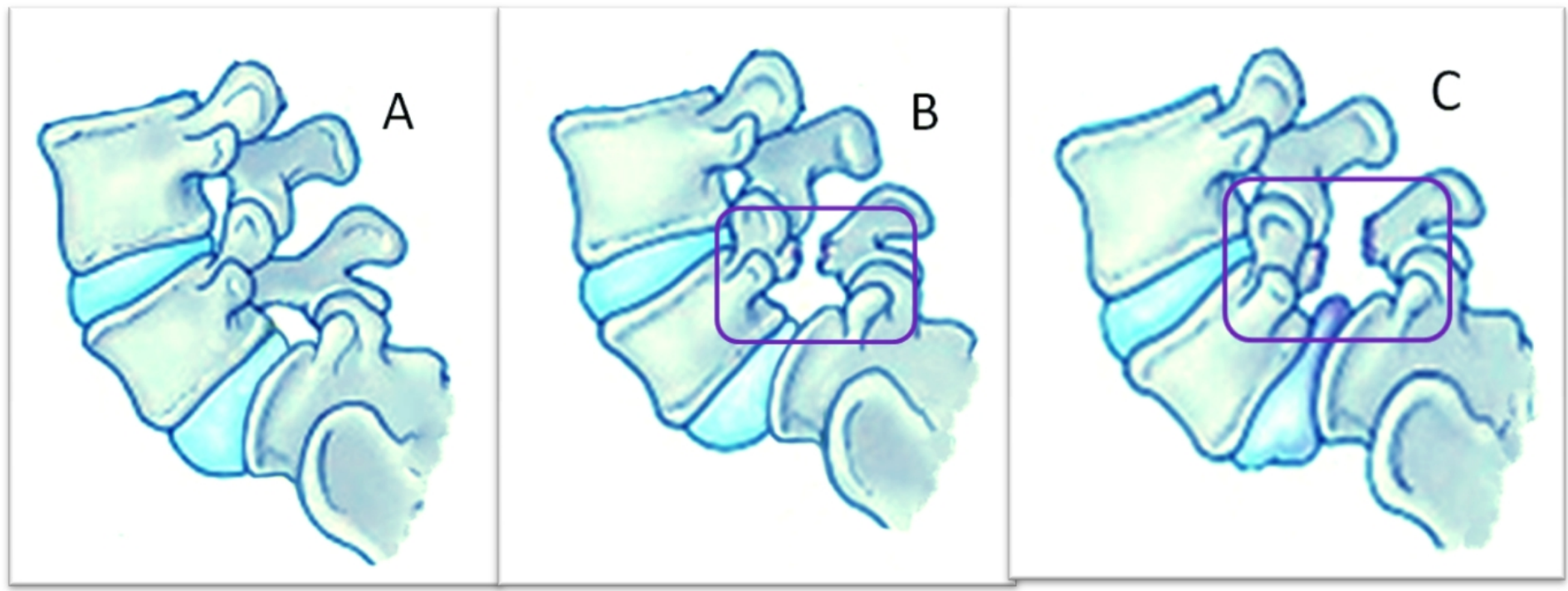

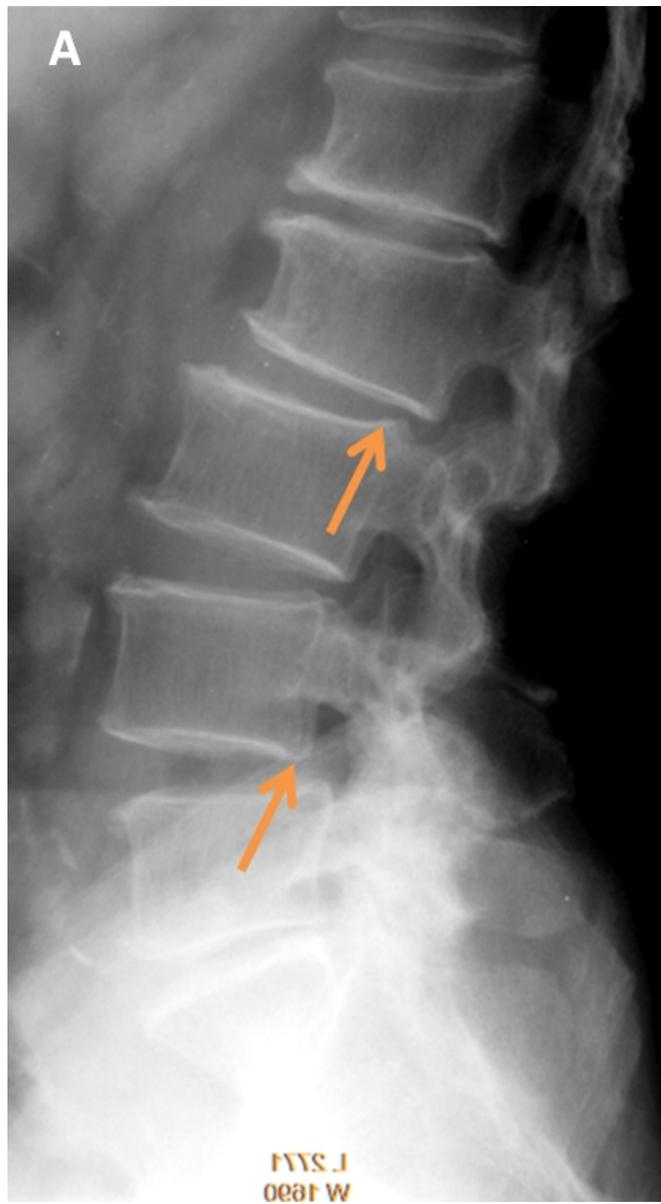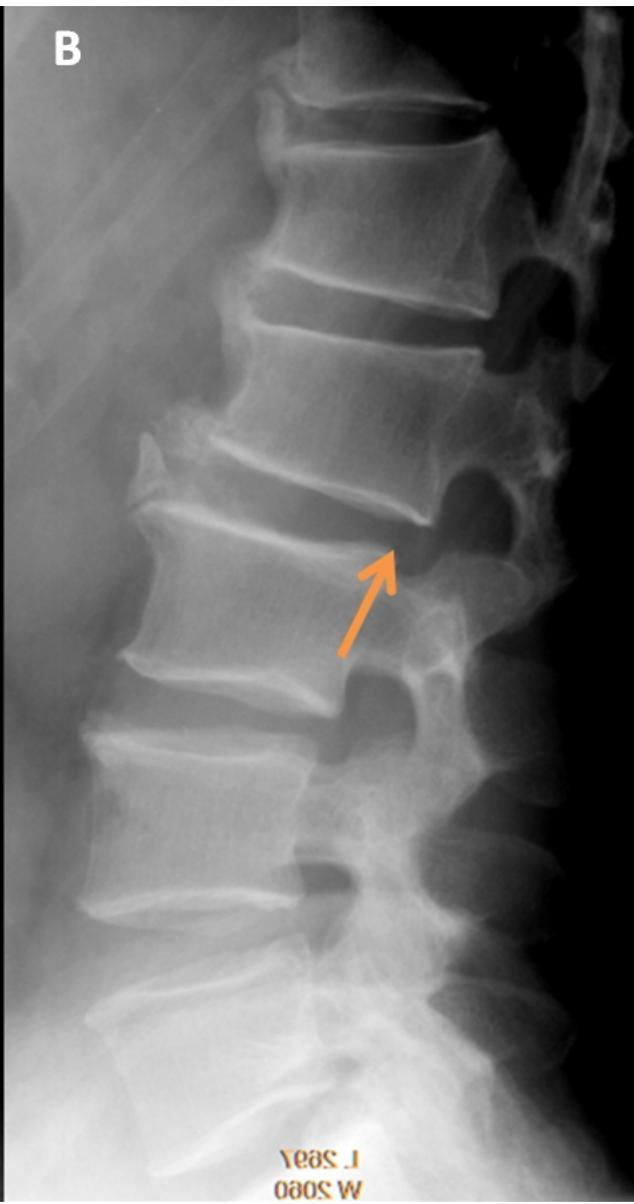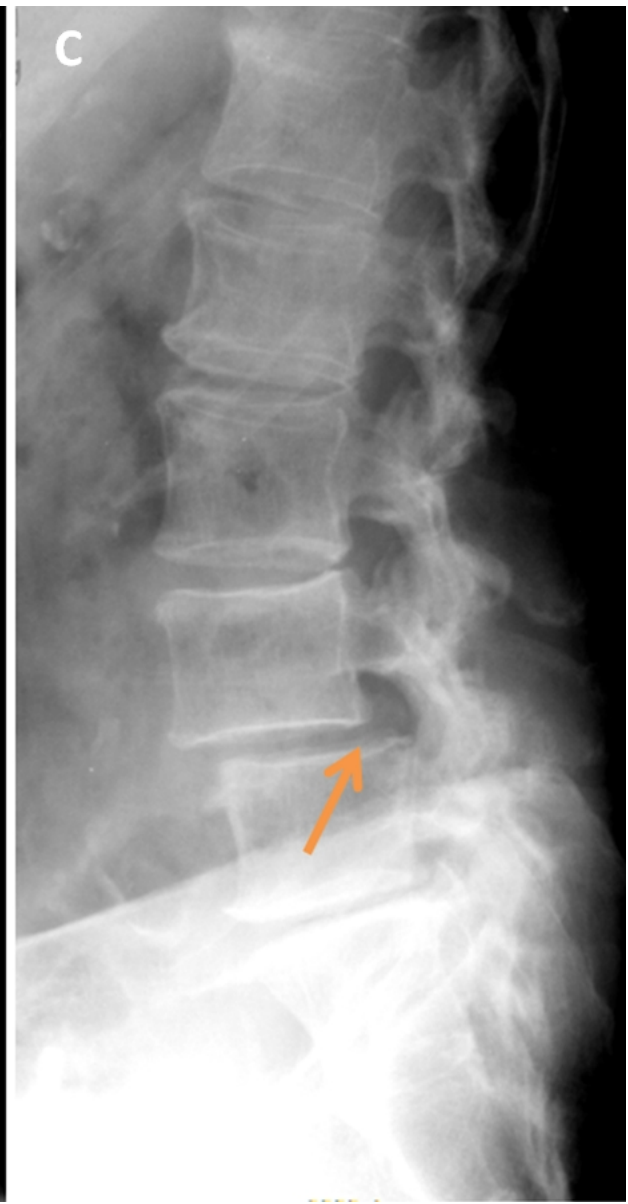

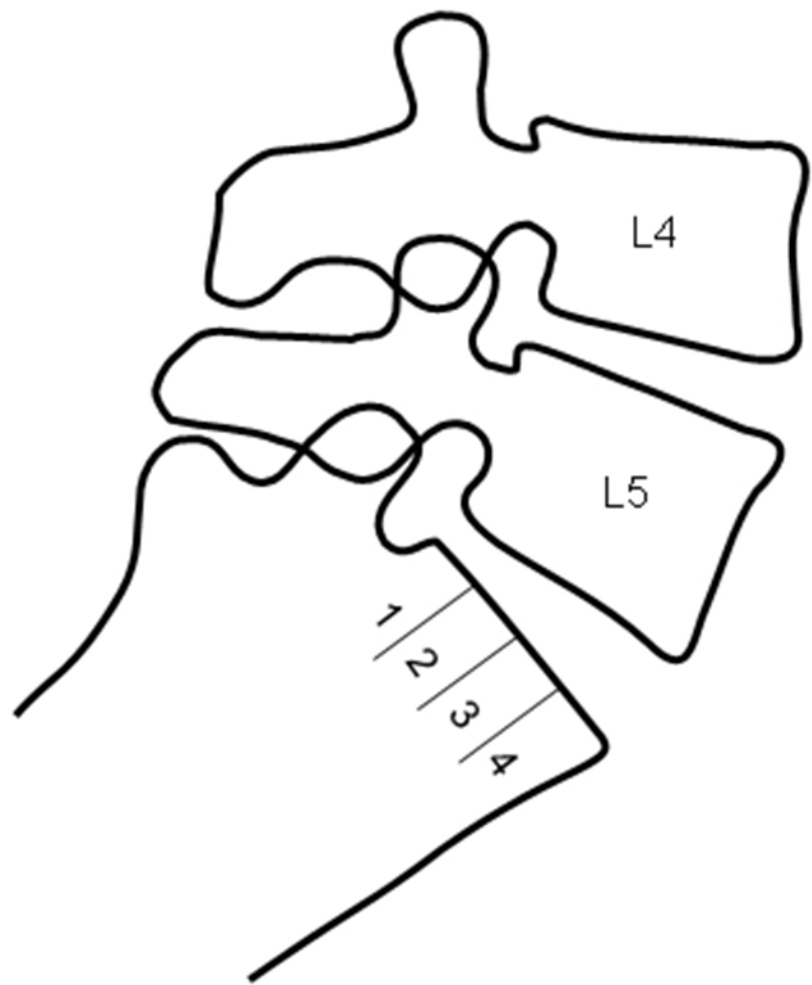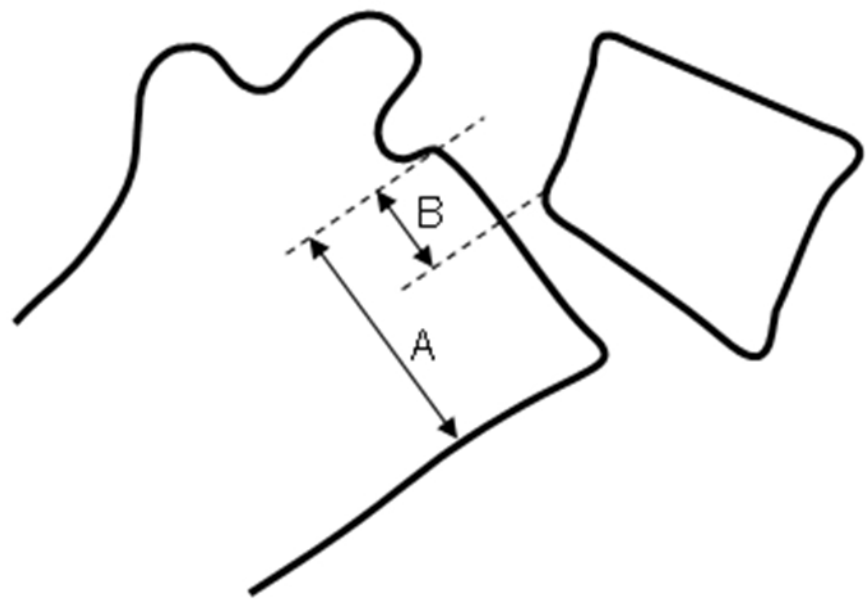

B/A x 100 = % of slip

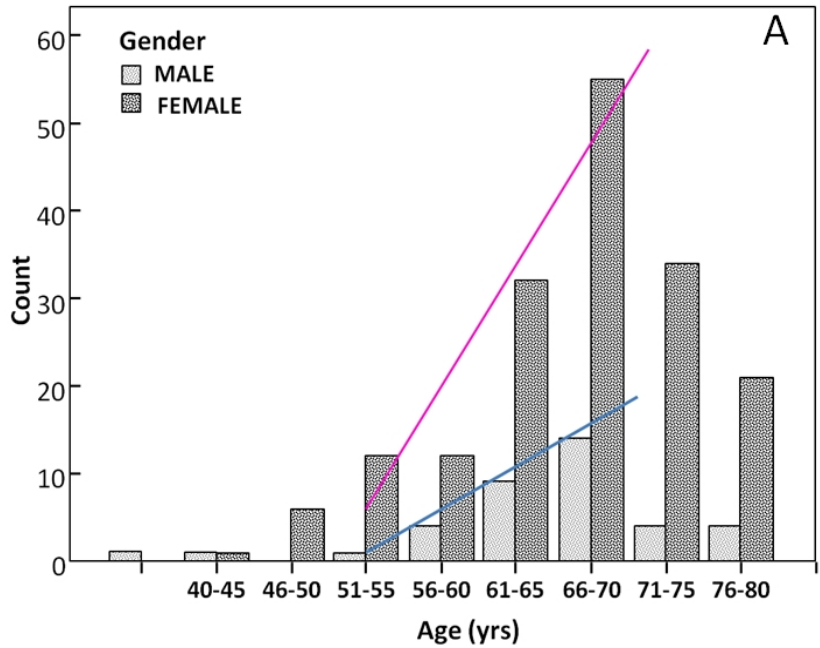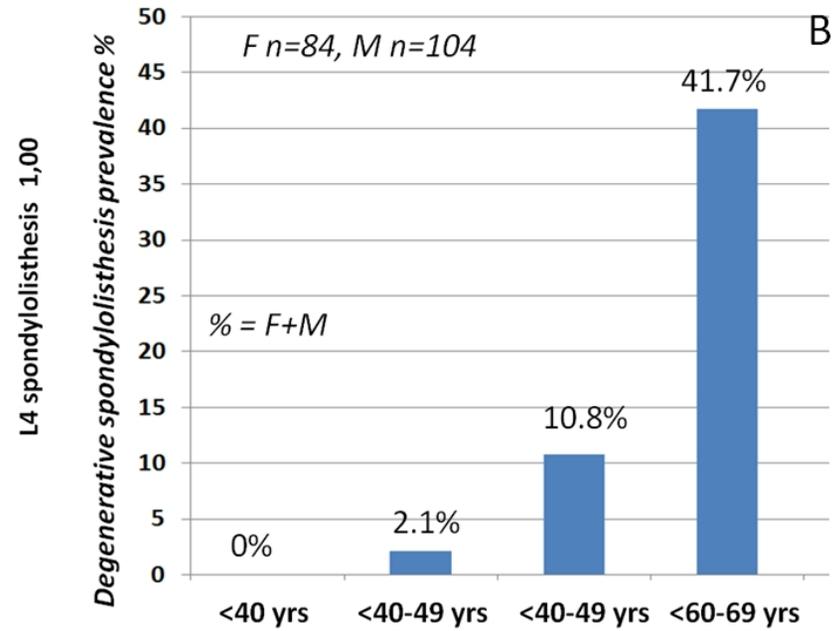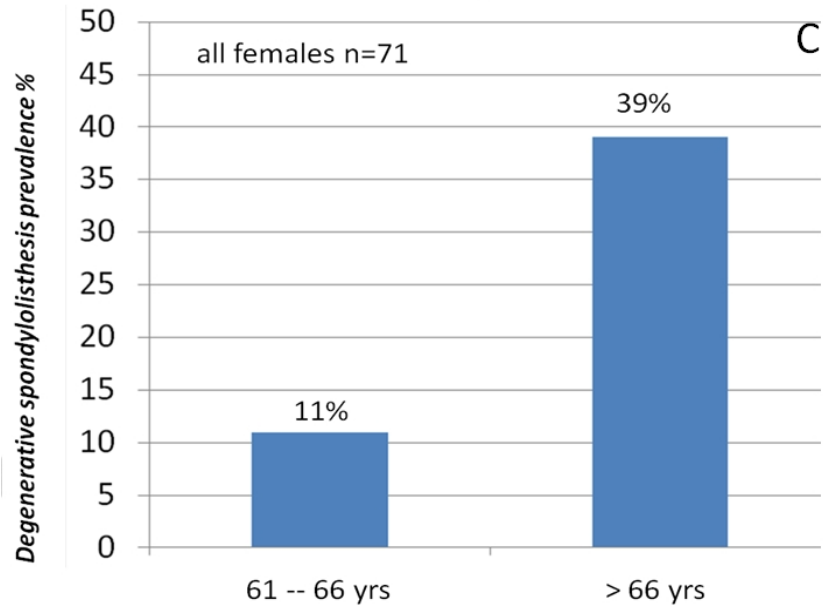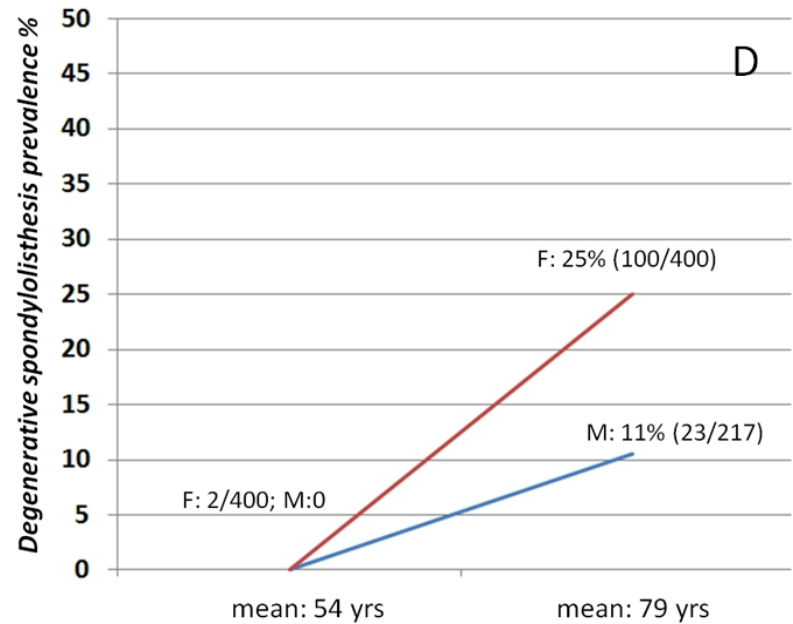

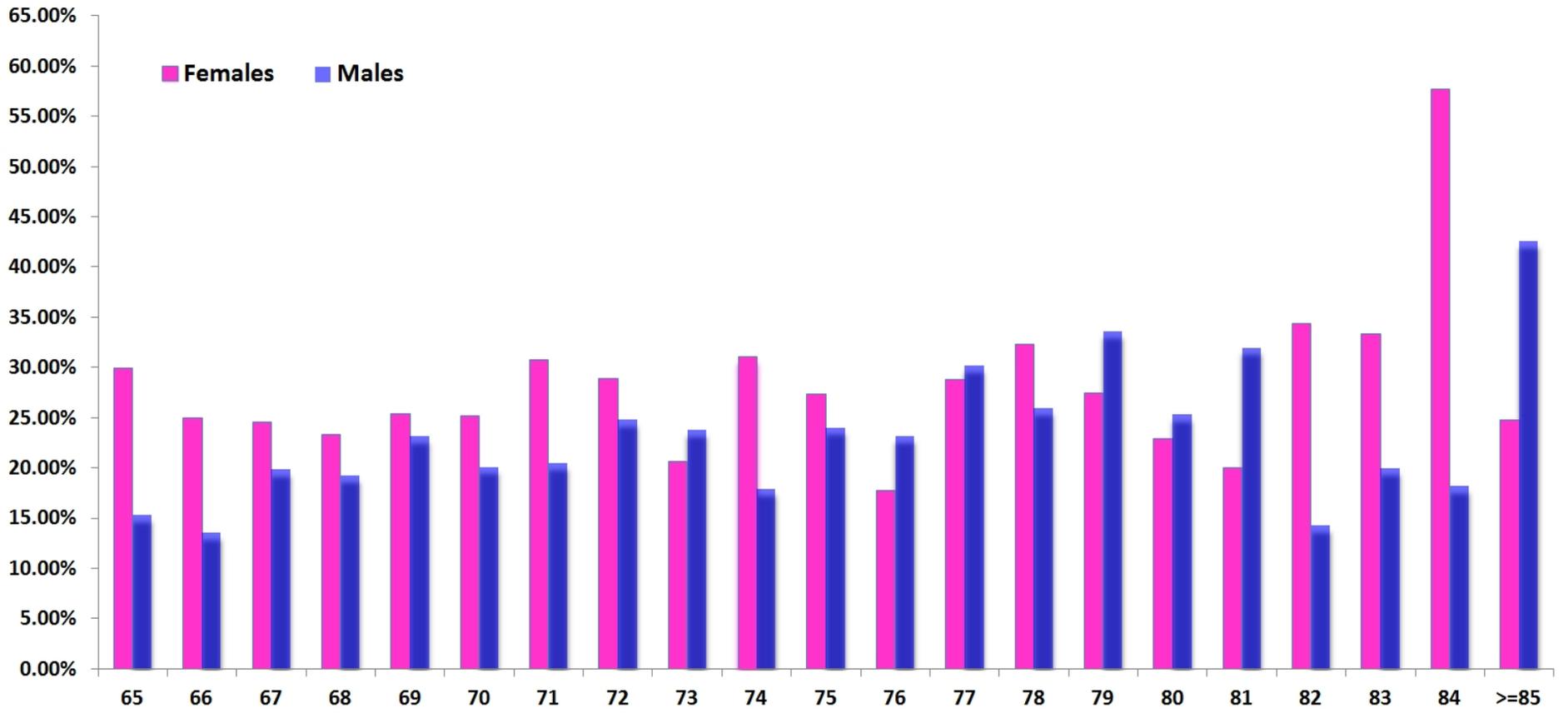

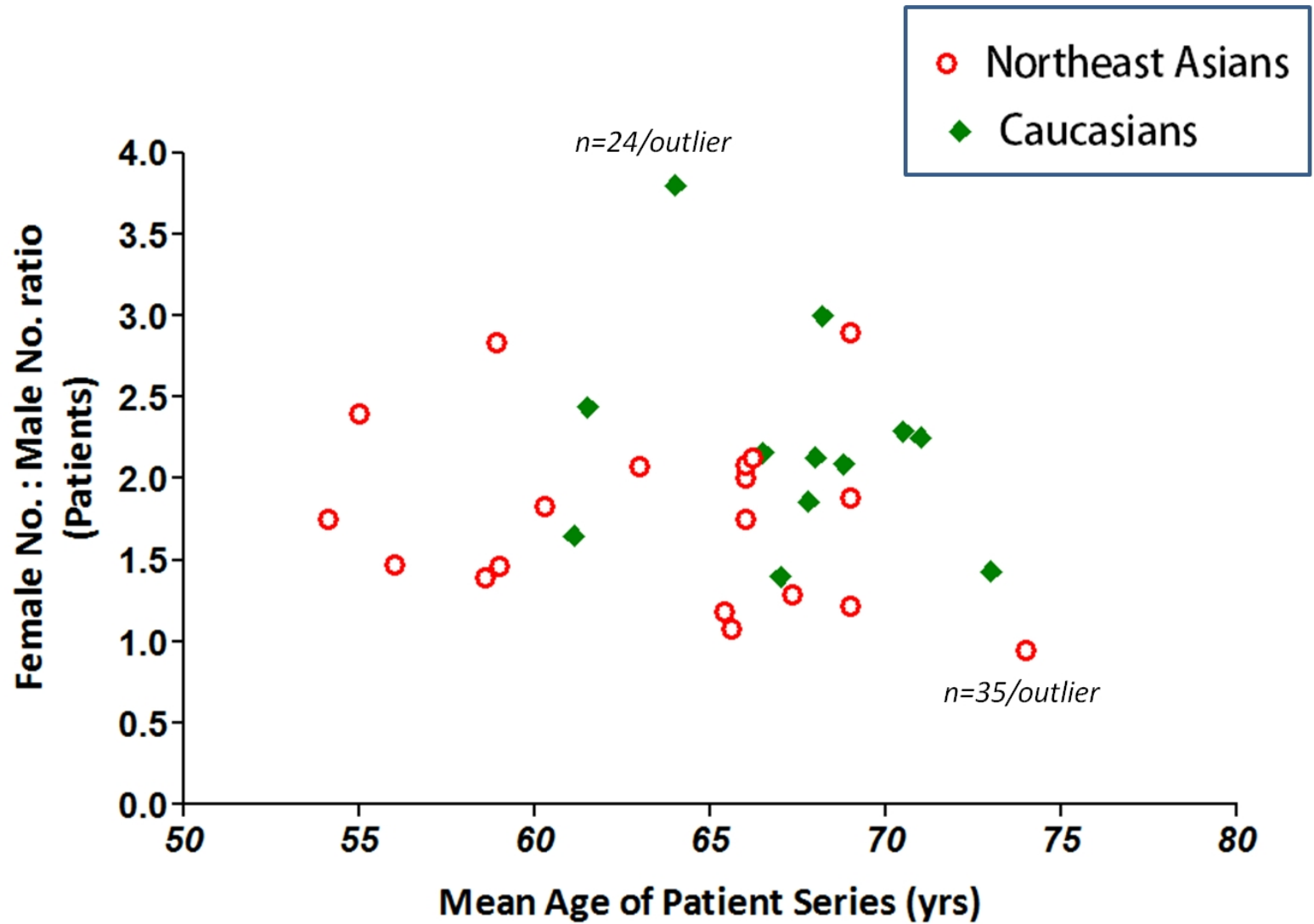

Table 1: Classification Systems for Spondylolisthesis

**Wiltse-Newman**
I. Dysplastic
II. Isthmic
    IIA, Disruption of pars as a result of stress fracture
    IIB, Elongation of pars without disruption related to repeated, healed microfractures
    IIC, Acute fracture through pars
III. Degenerative
IV. Traumatic
V. Pathologic

**Marchetti-Bartolozzi**
Developmental
    High dysplastic
        With lysis
        With elongation
    Low dysplastic
        With lysis
        With elongation
Acquired
    Traumatic
        Acute fracture
        Stress fracture
    Postsurgery
        Direct surgery
        Indirect surgery
    Pathologic
        Local pathology
        Systemic pathology
    Degenerative
        Primary
        Secondary

Lumbar degenerative spondylolisthesis epidemiology: a systemic review with a focus on gender-specific and age-specific prevalence

Yi Xiang J Wang[1], Zoltán Káplár[1], Min Deng[1], Jason C. S. Leung[2]

[1] Department of Imaging and Interventional Radiology, Faculty of Medicine, The Chinese University of Hong Kong, Prince of Wales Hospital, Shatin, New Territories, Hong Kong Specific Administrative Region.

[2] School of Public Health and Primary Care, Faculty of Medicine, The Chinese University of Hong Kong, Prince of Wales Hospital, Shatin, New Territories, Hong Kong Specific Administrative Region.

Correspondence to

Dr Yi Xiang J Wang, Department of Imaging and Interventional Radiology, Faculty of Medicine, The Chinese University of Hong Kong, Prince of Wales Hospital, Shatin, New Territories, Hong Kong Specific Administrative Region.

E-mail: yixiang_wang@cuhk.edu.hk  Tel (852) 2632 2289 Fax (852) 2636 0012

Running title: Lumbar degenerative spondylolisthesis epidemiology